# Expected values in percentile indicators

Lutz Bornmann* & Robin Haunschild**

*First author and corresponding author:

Division for Science and Innovation Studies

Administrative Headquarters of the Max Planck Society

Hofgartenstr. 8,

80539 Munich, Germany.

E-mail: bornmann@gv.mpg.de

**Max Planck Institute for Solid State Research

Heisenbergstr. 1,

70569 Stuttgart, Germany.

Email: R.Haunschild@fkf.mpg.de

**Abstract**

PP(top x%) is the proportion of papers of a unit (e.g. an institution or a group of researchers), which belongs to the x% most frequently cited papers in the corresponding fields and publication years. It has been proposed that x% of papers can be expected which belongs to the x% most frequently cited papers. In this Letter to the Editor we will present the results of an empirical test whether we can really have this expectation and how strong the deviations from the expected values are when many random samples are drawn from the database.

The Leiden Manifest presents ten guiding principles for research evaluation, especially for the proper use of bibliometrics in research evaluation. According to Hicks, Wouters, Waltman, de Rijcke, and Rafols (2015) "the most robust normalization method is based on percentiles: each paper is weighted on the basis of the percentile to which it belongs in the citation distribution of its field (the top 1%, 10% or 20%, for example)" (p. 430). PP(top x%) is the proportion of papers of a unit (e.g. an institution or a group of researchers), which belongs to the x% most frequently cited papers in the corresponding fields and publication years. The Leiden Ranking (http://www.leidenranking.com/ranking/2016/list) uses PP(top x%) as one of the central indicators to rank universities world-wide.

It is an important advantage of PP(top x%) that the indicator allows a comparison with an expected value. It has been proposed that x% of papers can be expected which belongs to the x% most frequently cited papers (e.g. Bornmann, Mutz, Marx, Schier, & Daniel, 2011). In this Letter to the Editor we will present the results of an empirical test whether we can really have this expectation and how strong the deviations from the expected values are when many random samples are drawn from the database.

The bibliometric data used in this paper are from an in-house database developed and maintained by the Max Planck Digital Library (MPDL, Munich) and derived from the Science Citation Index Expanded (SCI-E), Social Sciences Citation Index (SSCI), Arts and Humanities Citation Index (AHCI) prepared by Thomson Reuters (Philadelphia, Pennsylvania, USA). The in-house database contains not only bibliographic and times cited information for single papers, but also several field-normalized indicators. Three indicators are PP(top 50%), PP(top 10%), and PP(top 1%) which are calculated following Waltman and Schreiber (2013). These indicators consider ties in citation data if these ties are at the threshold separating the top papers from the bottom (100-x)%. The values of PP(top 50%), PP(top 10%), and PP(top 1%) for all papers published between 1980 and 2010 (n= 23154624) are PP(top 50%)=49.380, PP(top 10%)=9.904, and PP(top 1%)=0.990. The values are not exactly 50%, 10% and 1%, respectively, because the impact of the papers in our database is not fractionally assigned to subject categories. Instead, an average citation impact is calculated for papers assigned to more than one subject category. Waltman, van Eck, van Leeuwen, Visser, and van Raan (2011) explain with vivid examples how these deviations emerge if the impact is not fractionally measured.

Table 1. Key figures for PP(top 50%), PP(top 10%), and PP(top 1%) from 1000 random samples of different size

|         | Minimum | 1. quartile | Median | Mean  | 3. quartile | Maximum |
|---------|---------|-------------|--------|-------|-------------|---------|
| PP(top 50%) | | | | | | |
| 100     | 33.870  | 46.440      | 49.550 | 49.410 | 52.620     | 62.810  |
| 500     | 43.000  | 47.950      | 49.380 | 49.390 | 50.860     | 56.160  |
| 1000    | 45.350  | 48.410      | 49.410 | 49.410 | 50.440     | 53.850  |
| 10000   | 47.840  | 49.080      | 49.390 | 49.380 | 49.690     | 50.580  |
| 100000  | 48.920  | 49.270      | 49.370 | 49.370 | 49.480     | 49.900  |
| 1000000 | 49.210  | 49.350      | 49.380 | 49.380 | 49.410     | 49.510  |
| PP(top 10%) | | | | | | |
| 100     | 2.385   | 8.035       | 9.871  | 9.980 | 12.000      | 18.588  |
| 500     | 6.310   | 9.020       | 9.923  | 9.931 | 10.808      | 13.407  |
| 1000    | 6.591   | 9.249       | 9.887  | 9.870 | 10.454      | 12.835  |
| 10000   | 9.101   | 9.722       | 9.897  | 9.901 | 10.081      | 10.727  |
| 100000  | 9.564   | 9.847       | 9.899  | 9.903 | 9.965       | 10.163  |
| 1000000 | 9.804   | 9.885       | 9.904  | 9.904 | 9.923       | 9.981   |
| PP(top 1%) | | | | | | |
| 100     | 0.000   | 0.000       | 1.000  | 0.989 | 1.500       | 6.303   |
| 500     | 0.000   | 0.700       | 0.967  | 0.996 | 1.288       | 2.653   |
| 1000    | 0.200   | 0.773       | 0.977  | 0.986 | 1.176       | 1.977   |
| 10000   | 0.693   | 0.921       | 0.987  | 0.988 | 1.053       | 1.364   |
| 100000  | 0.892   | 0.970       | 0.992  | 0.991 | 1.011       | 1.094   |
| 1000000 | 0.964   | 0.984       | 0.990  | 0.990 | 0.997       | 1.031   |

Table 1 shows the key figures for PP(top 50%), PP(top 10%), and PP(top 1%) from 1000 random samples of different size. For example, if a random sample consisting of 500 papers has been drawn 1000 times from PP(top 10%) in the database (i.e. the papers of the population), the mean value is 9.931 which is very close to the population value of PP(top 10%)=9.903. However, the minimum and maximum are with PP(top 10%)=6.310 and PP(top 10%)=13.407 significantly lower and higher, respectively, than the population value. Similar results are visible for the other sample sizes and the analyzes based on PP(top 50%) and PP(top 1%). As expected, the deviations from the expected values of PP(top 50%)=49.380, PP(top 10%)=9.903, and PP(top 1%)=0.990, respectively, are larger, if the sample sizes become smaller (see Table 1).

The results of this small test based on an in-house database indicate two important things: (1) The expected value – i.e. the population value of a certain database – can be different from the expected value which results from the definition of the indicator. Thus, the population value of a database should be known, if the results of an empirical study (e.g. the bibliometric analysis of universities) based on a specific database are interpreted (using an expected value). (2) Although field-normalized indicators, like PP(top 50%), PP(top 10%), and PP(top 1%), are based on complex cross-field calculations, the expected values which are fixed by the indicators – 50%, 10%, and 1% - can really be expected when random samples are drawn multiple times or the sample sizes are large enough.